\documentclass[pra,floatfix,10pt]{revtex4}
\usepackage{color} 
\usepackage{epsfig}
\usepackage{amsmath}
\usepackage{graphicx}

\newcommand{\ene}{{\cal K}}
\newcommand{\eneq}{{\cal Q}}
\begin{document}
\title
{From Klein to anti-Klein tunneling in graphene tuning the Rashba spin-orbit interaction or the bilayer coupling}
\author{L. Dell'Anna$^1$, P. Majari$^2$, M.\,R. Setare$^2$}

\affiliation{
$^1$ Department of Physics and Astronomy, University of Padova, Italy\\
$^2$ Department of Science, University of Kurdistan, Sanandaj, Iran}

\date{\small{}}
\begin{abstract}
We calculate the transmission coefficient for a particle crossing a potential barrier in 
monolayer graphene with Rashba spin-orbit coupling and in bilayer graphene. 
We show that in both the cases one can go from Klein tunneling regime, 
characterized by perfect normal transmission, to anti-Klein tunneling regime, 
with perfect normal reflection, 
by tuning the Rashba spin-orbit coupling for a monolayer 
or the interplane coupling for a bilayer graphene. We show that 
the intermediate regime is characterized by a non-monotonic behavior 
\textcolor{black}{with oscillations and resonances in the normal transmission amplitude as a function of the coupling and of the potential parameters.} 
\end{abstract}
\maketitle
\section{Introduction}
Since its discovery \cite{novoselov2004}, graphene, the youngest of the carbon 
allotropes, a single layer of carbon atoms arranged in a hexagonal lattice,  
has attracted a lot of interest due to its novel and peculiar 
electronic transport properties \cite{castroneto2009}. 
Particles in graphene move according to a linear spectrum which leads to 
relativistic description of their dynamics \cite{novoselov2004},  
therefore, exhibit phenomena associated with relativistic fermions, such as the unusual half-integer quantum Hall effect \cite{peres2006} and the 
Klein tunneling \cite{katsnelson2006,cheianov2006,shytov2008}.
 
Its bilayer version, is made by two coupled monolayers of carbon atoms \cite{wang2017}. Many of the special properties of  bilayer graphene are similar to the 
monolayer one, such as  excellent electrical conductivity with room 
temperature, high thermal conductivity, strength and flexibility \cite{castroneto2009, ghosh2010, neekamal2010}. 
However the low-energy spectra of the  Dirac fermions in a monolayer and in a 
bilayer graphene are different \cite{mccann2007,wang2017bis}. Moreover, the 
tunneling in monolayer graphene is characterized by a perfect transmission 
of normal incident massless Dirac fermions through a potential barrier 
regardless of their hight and width \cite{katsnelson2007}, which is a consequence 
of the lack of backscattering due to conservation of the helicity. 
To prevent Klein effect, magnetic barriers have been proposed in order to 
 confine the quasiparticles in monolayer graphene \cite{demartino2007}. 
However, in the bilayer graphene  the incident and reflected states have the same pseudospin, while the transmitted state has the opposite one and due to the conservation of pseudospin, no Klein tunneling is expected.  
As a result, full reflection, known as anti-Klein tunneling \cite{katsnelson2006,kleptsyn2015,varlet2016} occurs. 
As we will see, an analogous result occurs for monolayer graphene in the presence of strong Rashba spin-orbit coupling \cite{liu2012}. 
Rashba and intrinsic spin-orbit couplings (SOC) are two types of spin-orbit interactions that can be present in graphene \cite{kane2005}. 
The moving of electrons in the atomic electric field produce a magnetic field 
which interacts with the electron spin, generating a weak intrinsic SOC. However, the extrinsic Rashba 
spin-orbit coupling term arises due to an external electric field 
perpendicular to the graphene sheet. We must say that the nearest-neighbor 
intrinsic SOC vanishes due to the symmetry of the graphene lattice while the 
next-nearest-neighbor coupling is nonzero \cite{kochan2017}. 
On the  other hand, Rashba effect appears in graphene 
if the mirror symmetry of the system is broken and causes a 
nearest-neighbor extrinsic SOC \cite{bychkov1984}. 
Here, we will focus on the Rashba SOC 
and neglect the intrinsic one \cite{qiao2013,min2006}.
We will see that at low energy limit the single-layer graphene with  
Rashba spin-orbit coupling is described by the same Hamiltonian of a bilayer 
graphene with the interlayer coupling which plays the role of SOC in the other system. This observation will allow us to treat the two cases at once. In both 
cases, indeed, we will show how one can approach anti-Klein tunneling 
by increasing the two types of couplings in the two different setups. 

\section{The two models}
Here we will briefly present the Hamiltonian describing the dynamics of the quasiparticles in a monolayer graphene in the presence of spin-orbit coupling and
the Hamiltonian related to two layers of graphene with interlayer coupling. 
We will show that the two models are described by two Hamiltonians which can be mapped one to the other by a simple unitary transformation.
\subsection{Monolayer graphene with spin-orbit coupling}

We consider a tight-binding model including Rashba spin-orbit coupling 
and study the scattering problem in single-layer graphene. In the presence of  
Rashba and intrinsic spin-orbit coupling, the Hamiltonian of monolayer 
graphene is written as
\begin{equation}\label{1}
H=H_0+H_R+H_{SO}.
\end{equation}
$H_0$ describes the noninteracting Hamiltonian which is given by
\begin{equation}\label{2}
H_0=-t\sum_{<i,j>,\alpha}c_{i\alpha}^{\dag}c_{j\alpha} ,
\end{equation}
where $c_{i\alpha}^{\dag}$ and  $c_{j\alpha}$ represent the creation and annihilation operators for electrons placed on the site $i$, while $\alpha$ labels the spin and  \textcolor{black}{$t\approx 2.8~eV$ \cite{castroneto2009}} is the intralayer hopping parameter between nearest-neighbor sites. The Rashba Hamiltonian is also given by a nearest-neighbor 
hopping term \cite{qiao2012}
\begin{equation}\label{3}
H_R=it_R\sum_{<i,j>} c_{i}^{\dag}\hat{e}_z .(\vec{s}\times \vec{d}_{ij})c_{j} ,
\end{equation}
where $\vec{d}_{ij}$ is a lattice vector pointing from
site $j$ to site $i$, 
\textcolor{black}{$\vec{s}$ is a vector whose elements are the Pauli matrices in the spin space,}  
the spin-orbit coupling $t_R$ is determined 
by the strength of the electric field, and 
$c^\dagger_i=(c^\dagger_{i\uparrow},c^\dagger_{i\downarrow})$ 
are vectors in the spin space.
 We can write intrinsic spin-orbit coupling (SOC) Hamiltonian as
\begin{equation}\label{4}
H_{SO}={2i\over \sqrt{3}}t_{SO}\sum_{\ll i,j\gg}c_{i}^{\dag}\, \vec{s} .(\vec{d}_{kj}\times \vec{d}_{ik})c_{j} ,
\end{equation}
where $\ll .\gg$ means that we sum up to next-nearest neighbour lattice sites. 
Expanding the tight-binding Hamiltonian Eq.({\ref {1}}) in the vicinity 
of the valleys $\vec{K}_{\pm}=(\pm 4\pi/3\sqrt{3},0)$, gives the
following low-energy Hamiltonian in the sublattice, $(A,B)$, and spin, 
$(\uparrow,\downarrow)$, spaces 
\begin{equation}\label{5}
{\cal H}=v(\eta \sigma_x p_x+\sigma_y p_y)
+{\lambda_R}(\eta {\sigma}_x 
s_y -{\sigma}_y 
{s_x})+\eta \lambda_{o} \sigma_z
s_z,
\end{equation}
where $\eta=\pm 1$ labels the valley degrees of freedom, $\sigma_x$, $\sigma_y$, $\sigma_z$ are Pauli matrices acting on the pseudospin (or sublattice) 
space, and $v={3\over 2}ta$ ($a$ the lattice spacing) is the Fermi velocity. 
The estimated values of  $ \lambda_{o}$  and $\lambda_R$ 
remain rather controversial, however, by tight-binding calculation, 
the value of intrinsic SOC is found to be $ \lambda_{o}=3 \sqrt{3} t_{SO}= 12 \mu eV$ \cite{konschuh2010}, while the valuation of the Rashba coupling is $\lambda_R ={3t_R\over 2}= 37.4 \mu eV $ \cite{ast2012}. 
However, experimentally the Rashba coupling can be strongly 
enhanced by appropriate optimization of the substrate 
up to values of the order of $\lambda_R=14 meV$ \cite{varykhalov}. 
For that reason we will consider only the Rashba spin-orbit interaction 
$\lambda_R$, neglecting $\lambda_{o}$, and 
rewriting the low-energy  Hamitoniam around a single Dirac point 
(for instance, $\eta=+1$) on the double-spinor basis 
$\Psi({\bf r})=(\psi_{A \uparrow}({\bf r}), \psi_{B \downarrow}({\bf r}), \psi_{B \uparrow}({\bf r}), \psi_{A \downarrow}({\bf r}))^t$, 
such that $H=\int d_{\bf r}\Psi^\dagger({\bf r}){\cal H}\,\Psi({\bf r})$, with 
\begin{equation}\label{HS}
{\cal H}=
\begin{pmatrix}
0 & 0 &  v(p_x-ip_y)  & 0\\
0 & 0 & 0 &  v(p_x+ip_y) \\
v(p_x+ip_y) & 0 & 0 & {-i\lambda} \\
0& v(p_x-ip_y) &  {i\lambda } & 0
\end{pmatrix}
=
\begin{pmatrix}
{\cal H}_{1} &  {T}\\
{T}^* & {\cal H}_{2}
\end{pmatrix}
\end{equation}
where, for the sake of simplicity, 
$\lambda=2\lambda_R$. Dividing the four-bands Hamitoniam 
into $2\times 2$ blocks and solving the eigenvalue equation for energy $E$ we can obtain the reduced effective two-bands Hamiltonian acting on the spinor $(\psi_{A\uparrow}({\bf r}),\psi_{B\downarrow}({\bf r}))^t$ 
\begin{equation}
\label{He}
{\cal H}_{e}\equiv {\cal H}_1+T(E-{\cal H}_2)^{-1} T^*,
\end{equation}
which, 
in the vicinity of the Dirac point, namely for $E\rightarrow 0$, becomes 
${\cal H}_{e}\simeq {\cal H}_1-T{\cal H}_2^{-1} T^*$, 
the leading terms in the perturbative expansion in $E$, therefore
\begin{equation}\label{HeffS}
{\cal H}_{e}\simeq -\begin{pmatrix}
0 &  {-i {v^2 \over{ \lambda }}(p_x-ip_y)^2}\\
{i {v^2 \over{ \lambda }}(p_x+ip_y)^2} & 0
\end{pmatrix}.
\end{equation}
Notice that, after rotating the spinor along the quantization axis by $\pi/4$, 
namely $U {\cal H}_e U^\dagger$, with $U=e^{i\sigma_z\pi/4}$ (in the reduced 
two-bands model the spin and the pseudospin are locked),   
one ends up to the analogous effective Hamiltonian for a bilayer graphene, 
see Eq.~(\ref{HeffB}).  
As it is known \cite{katsnelson2006} for bilayer graphene, 
under certain conditions, perfect normal reflection occurs, 
therefore also for single-layer graphene with strong spin-orbit 
coupling we expect the same behavior, as we will see in what follows.

\subsection{Bilayer graphene}
\label{sec.bilayer}
Bilayer graphene  consists of two layers of graphene with  
inequivalent sites $A$ and $B$  in the top layer labeled by $1$ and $A$ and 
$B$ in the bottom layer labeled by $2$. These two layers can be arranged
according to Bernal stacking ($AB$-stacking) or $AA$-stacking 
\cite{varlet2016}. 
Here, we will consider  the Bernal stacking bilayer graphene. 
The low energy Hamiltonian at one valley, let us consider $\vec{K}_+$, 
choosing the basis 
$( \psi _{A1}({\bf r}), \psi _{B2}({\bf r}), \psi _{B1}({\bf r}), 
\psi _{A2}({\bf r}) )^t$, in order 
to make easier the comparison with the previous case, is  
given by 
%
\begin{equation}\label{HB}
{\cal H}=
\begin{pmatrix}
0 & 0 &  v(p_x-ip_y)  & 0\\
0 & 0 & 0 &  v(p_x+ip_y) \\
v(p_x+ip_y) & 0 & 0 & {\lambda } \\
0& v(p_x-ip_y) &  {\lambda } & 0
\end{pmatrix}
\end{equation}
where 
now $\lambda\approx 0.4\,eV$ is the interlayer 
hopping energy between atom $A2$ and atom $B1$,  
further skew hopping parameters are negligible \cite{duppen2013,masir2009}. 
Actually the interplane coupling between atoms $A2$ and $B1$ can be written 
in the Hamiltonian $H$ as a term $\lambda(\sigma_y \tau_y+\sigma_x \tau_x)/2$, where $\tau_{x,y,z}$ are Pauli matrices acting on the interlayer indices 
$(1,2)$, which play the role of real spins in the previous case. 
Identifing 
the pseudospin related to the interlayer space with the real spin, after a simple rotation $U=e^{-i\pi\tau_z/4}$, applied to the Hamiltonian, $U{\cal H}U^\dagger$, one goes from Eq.~(\ref{HB}) to Eq.~(\ref{HS}), or, viceversa, appling $U=e^{i\pi s_z/4}$ to Eq.~(\ref{HS}) one gets Eq.~(\ref{HB}).
Moreover, 
by the same procedure seen above, Eq.~(\ref{He}), 
it is possible to obtain an effective two-bands Hamiltonian for the 
components $(\psi _{A1}, \psi _{B2})$ 
in the limit of strong interlayer coupling $\lambda$, 
which 
is given by \cite{mucha2010,tudorovskiy, novoselov2006}
\begin{equation}\label{HeffB}
{\cal H}_{e}\simeq -\begin{pmatrix}
0 &  {v^2 \over{\lambda }}(p_x-ip_y)^2\\
{v^2 \over{ \lambda }}(p_x+ip_y)^2 & 0
\end{pmatrix}.
\end{equation}
In this reduced two-bands model the sublattice and interlayer pseudospins 
are locked and, as already mentioned, Eq.~(\ref{HeffS}) 
and Eq.~(\ref{HeffB}) are mapped one to the other by a $\pi/4$-rotation around 
the quantization axis of the two-component spinor.  
While the two layers of graphene, when separated, exhibit Klein effect, in the limit of strong interlayer coupling, using the reduced Hamiltonian 
Eq.~(\ref{HeffB}), where $\lambda/2v^2$ is the so-called effective mass, 
anti-Klein tunneling occurs \cite{katsnelson2006}. 


\section{Tunneling through a potential barrier}
  
Since the two models seen above are described by two Hamlitonians which are 
related by a rotation along the spin (or interlayer pseudospin) quantization 
axis, 
we can treat both the cases at the same time and we will show that 
the tunneling properties are the same. The aim of the work 
is to show how to go from Klein tunneling,  
peculiar for monolayer graphene, to anti-Klein tunneling occurring in bilayer 
or \textcolor{black}{in monolayer graphene with} 
strong Rashba spin-orbit coupling.  
We will consider, therefore, the scattering problem of a charge carrier 
through a scalar potential barrier \textcolor{black}{with width $d$}, 
for a generic coupling $\lambda$ 
(Rashba coupling or interlayer coupling) 
when the energy of the incident particle is smaller than the 
height of the potential ($E< V_0$) 
\begin{equation}\label{Vx}
V(x) =
\begin{cases}
0\,,    \; \; \textrm{for} \; \;  x< {0} \\
 V_0\,,    \; \textrm{for} \; \; 0\leq x \leq d \\
 0\,,    \; \; \textrm{for} \; \;  x> {d}
\end{cases}
\end{equation}
Before proceding with the solution of the scattering problem, let us 
write the eigenvalues and eigenstates of the two Hamiltonians in 
Eqs. (\ref{HS}) and (\ref{HB}). In both cases 
the four eigenvalues are 
$E=\pm\frac{1}{2}(\lambda\pm\sqrt{\lambda^2+4v^2(k_x^2+k_y^2})$, 
where $k_{x,y}$ are the eigenvalues of the operators $p_{x,y}$ (we put $\hbar=1$.) 
The eigenstates of Eq.~(\ref{HS}) are $e^{i y k_y }$ times 
the columns of the following matrix 

\begin{equation}\label{W}
{\cal W}_E(x)=
\begin{pmatrix}
e^{ixk_x^{(1)} } & e^{-ixk_x^{(1)} } & e^{ixk_x^{(2)} }  & e^{-ix k_x^{(2)}}\\
\frac{-k_y+ik_x^{(1)}}{ik_y-k_x^{(1)}}\, e^{ixk_x^{(1)} } &
\frac{-k_y-ik_x^{(1)}}{ik_y+k_x^{(1)}}\, e^{-ixk_x^{(1)} }&
\frac{k_y-ik_x^{(2)}}{ik_y-k_x^{(2)}}\, e^{ixk_x^{(2)} }&
\frac{k_y+ik_x^{(2)}}{ik_y+k_x^{(2)}}\, e^{-ixk_x^{(2)} }\\
\frac{-E/v}{ik_y-k_x^{(1)}}\, e^{ixk_x^{(1)} } &
\frac{-E/v}{ik_y+k_x^{(1)}}\, e^{-ixk_x^{(1)} }&
\frac{-E/v}{ik_y-k_x^{(2)}}\, e^{ixk_x^{(2)} }&
\frac{-E/v}{ik_y+k_x^{(2)}}\, e^{-ixk_x^{(2)} }\\
\frac{iE/v}{ik_y-k_x^{(1)}}\, e^{ixk_x^{(1)} } &
\frac{iE/v}{ik_y+k_x^{(1)}}\, e^{-ixk_x^{(1)} }&
\frac{-iE/v}{ik_y-k_x^{(2)}}\, e^{ixk_x^{(2)}}&
\frac{-iE/v}{ik_y+k_x^{(2)}}\, e^{-ixk_x^{(2)}}\\
\end{pmatrix}
\end{equation}
where
\begin{eqnarray}
\label{k1}
k_x^{(1)}=\frac{1}{v}\sqrt{E(E+\lambda)-v^2k_y^2}\\
k_x^{(2)}=\frac{1}{v}\sqrt{E(E-\lambda)-v^2k_y^2}
\label{k2}
\end{eqnarray}
For the bilayer case we have to replace Eq.~(\ref{W}) with the matrix which diagonalizes Eq.~(\ref{HB}), that can be written in terms of Eq.~(\ref{W}) as it follows
\begin{equation}
\label{Wb}
{\cal W}^b_{E}(x)=I_b{\cal W}_{E}(x)
\end{equation} 
where 
\begin{equation}\label{Ib}
I_b=
\begin{pmatrix}
1 & 0 & 0 & 0\\
0 & -i & 0 & 0\\
0 & 0 & 1 & 0\\
0 & 0 & 0 & -i
\end{pmatrix}
\end{equation}
so that, as we will see, the tunneling properties for the two cases are exactly the same.
\subsection{Transfer matrix approach}
In order to solve the Shr\"odinger equation in the presence of a 
picewise constant potential as in Eq.~(\ref{Vx}) one has to impose the continuity 
condition of the wavefunction at the boundaries $x=0$ and $x=d$ which separate 
region $I$ ($x<0$) where $V(x)=0$, region $II$ ($0\le x\le d$) where $V(x)=V_0$ and 
region $III$ ($x>d$) where $V(x)=0$. 
This results to writing, for any $j$, 
\begin{eqnarray}
&&\sum_{j=1}^4 {\cal W}_E(0)_{ij}\, c^I_{j}
=\sum_{j=1}^4 {\cal W}_{E-V_0}(0)_{ij}\, c^{II}_j\\
&&\sum_{j=1}^4 {\cal W}_{E-V_0}(d)_{ij}\, c^{II}_{j}
=\sum_{j=1}^4 {\cal W}_{E}(d)_{ij}\, c^{III}_j
\end{eqnarray}
where $c^I_i$, $c^{II}_i$, $c^{III}_i$ are some coefficients of the wavefunctions written in terms of the eigenstates of the Hamiltonian, defined on 
the three regions. Solving for the intermediate state one can write the wavefunction 
defined on the right region in terms of its value in the left region
\begin{equation}
c^{I}_i=\sum_{j=1}^4{\cal T}_{ij}c^{III}_j
\end{equation}
where we define the following transfer matrix
\begin{equation}
\label{transferM}
{\cal T}={\cal W}^{-1}_E(0)\,{\cal W}_{E-V_0}(0)\,
{\cal W}^{-1}_{E-V_0}(d)\,{\cal W}_{E}(d)
\end{equation}
which will encode the continuity of the wavefunction at the two 
interfaces, at $x=0$ and $x=d$. 
It is easy to see that this matrix is the same for both the Rashba case in a 
monolayer and in the bilayer, since, in the latter case, one should  
insert $I_b^{-1}I_b$ twice, at the two interfaces, but this quantity is an identity 
since $I_b$ does not depend on the energy. As a result, 
${\cal T}_b$, for the bilayer, obtained replacing   
${\cal W}$ with ${\cal W}^b$ 
from Eq.~(\ref{Wb}), is the same as in the monolayer case with Rashba coupling, 
${\cal T}_b={\cal T}$. 

\subsection{Results}
\textcolor{black}{Let us now apply this approach to find the transmission 
probability for a particle to go across the barrier}. 
We have to distinguish two cases according to the momentum carried by the 
incident particle. Let us 
consider, for positive energy, 
\emph{i)} a particle coming from the first (lower) band 
and \emph{ii)} a particle coming from the second (upper) band.  
\begin{figure}
\begin{center}
\includegraphics[width=.15\textwidth]{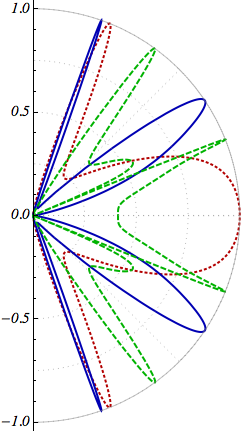}
\hspace{1cm}
\includegraphics[width=.15\textwidth]{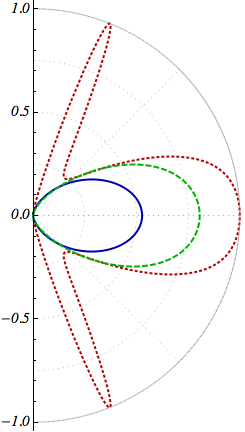}
\end{center}
\caption{Transmission probabilities 
for incident quasiparticles with positive energy $E=2$ trought a barrier 
of height $V_0=5$ and width $d=4$. Units of measure: fixing for $d$ the arbitrary 
unit length $\ell$, the energies $E$, $V_0$, $\lambda$ are expressed 
in unit of $\hbar v/\ell$. If 
we choose $\ell\sim 100\,nm$, the energy scale is $\hbar v/\ell\sim 5\, meV$, 
order of magnitude of the Rashba coupling. For $\ell\sim 1\,nm$, the energy scale is 
$\hbar v/\ell\sim 0.5\, eV$, same order of magnitude of the bilayer coupling.
(Left) Transmission probability $|t_1(\phi)|^2$ of a quasiparticle 
from first lower band 
with momentum $k_x^{(1)}$, Eq.~(\ref{k1}), as a 
function of the angle $\phi\in [-\pi/2,\pi/2]$ 
and different values of the coupling $\lambda$: 
$\lambda=0$ (red dotted line), $\lambda=1.2$ (green dashed line), $\lambda=2$ 
(blue solid line). 
(Right) Transmission probability $|t_2(\phi)|^2$ 
of aquasiparticle from second upper band
with momentum $k_x^{(2)}$, Eq.~(\ref{k2}), as a
function of the angle $\phi$ and different values of $\lambda$:
$\lambda=0$ (red dotted line), $\lambda=0.5$ (green dashed line), $\lambda=1$
(blue solid line).}
\label{transmission_angle}
\end{figure}
\subsubsection{Lower band}
Let us consider an incident particle traveling with momentum $k_x^{(1)}$, Eq.~(\ref{k1}). 
In this case the scattering problem can be formulated in terms of the 
following matrix equation
\begin{equation}
\label{eq1band}
\begin{pmatrix}
1\\
r_1\\
0\\
r_2
\end{pmatrix}=
{\cal T}
\begin{pmatrix}
t_1\\
0\\
t_2\\
0
\end{pmatrix}
\end{equation}
where $r_1$ is the reflection coefficient (for a wavefunction with momentum $-k_x^{(1)}$) and $t_1$ the transmission one. In principle one can allow for a 
reflected wave with momentum $-k_x^{(2)}$, therefore $r_2$, and a transmitted 
one ($t_2$) with momentum $k_x^{(2)}$.  
Defining
\begin{equation}
\ene_E=\frac{1}{v}\sqrt{E(E+\lambda)}
\end{equation}
the momentum can be written in terms of the incident angle $\phi$
\begin{eqnarray}
k_x=\ene_E \cos\phi\\
k_y=\ene_E \sin\phi
\label{kyphi}
\end{eqnarray}
Notice that, since $V(x)$ is only along $x$-direction, $k_y$ is the same 
everywhere, also inside region $II$, 
namely $k_y=\ene_E \sin\phi=\ene_{E-V_0} \sin\theta$, which defines 
the angle of refraction $\theta$ as for the Snell law in optics. \\
Using Eq.~(\ref{kyphi}) in Eq.~(\ref{k1}), one can notice that, 
for $\lambda>(V_0-E)$, the wave inside the barrier becomes evanescent
since $\ene_{E-V_0}=\frac{i}{v}\sqrt{|(E-V_0)(E-V_0+\lambda)|}$, therefore 
one can expect a suppression of the transmission for such values of the coupling.  
Now writing Eqs.~(\ref{W}), (\ref{transferM}) in terms of the angle of incidence $\phi$ one can solve Eq.~(\ref{eq1band}), getting in particular $t_1(\phi)$, the Fresnel 
transmission coefficient for an incident 
particle carrying momentum $(k_x, k_y)=\ene_E(\cos\phi,\sin\phi)$.  
In Fig.~\ref{transmission_angle} (left plot) examples of the transmission probability $|t_1(\phi)|^2$ for different values of $\lambda$ are reported. 
\begin{figure}[h!]
\includegraphics[width=.37\textwidth]{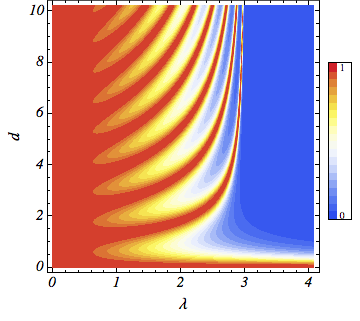}
\caption{Transmission probability $|t_1(0)|^2$, from Eq.~(\ref{t1_0}), 
for a normally incident quasiparticle coming from 
the first lower band with energy $E=2$ trought a barrier
of height $V_0=5$ (both in unit of $\hbar v/\ell$), with momentum 
$k_x^{(1)}$, Eq.~(\ref{k1}), 
as a function of the width of the barrier $d$ (in unit of $\ell$, 
arbitrary length scale) and strength 
of the coupling
$\lambda$ (in unit of $\hbar v/\ell$). For $\lambda>V_0-E$ the normal transmission is strongly suppressed.}
\label{fig.cont1}
\end{figure}
\begin{figure}[h!]
\includegraphics[width=.37\textwidth]{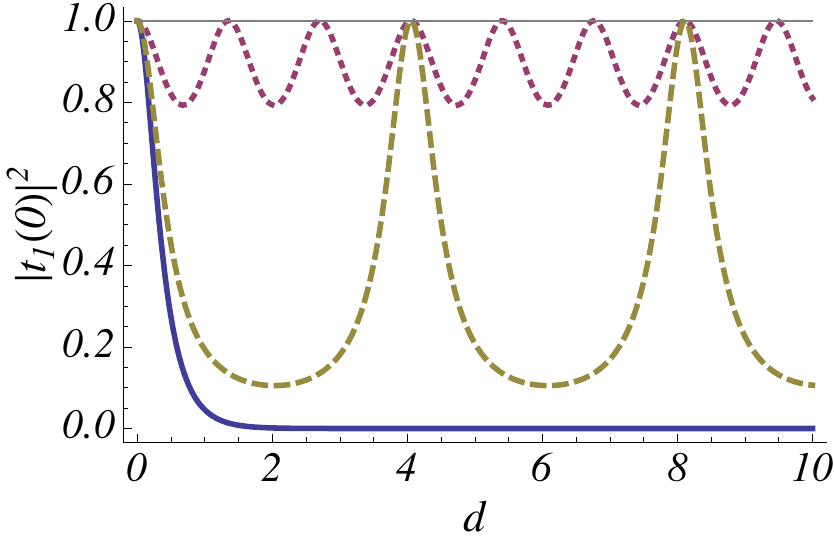}\hspace{0.3cm}
\includegraphics[width=.37\textwidth]{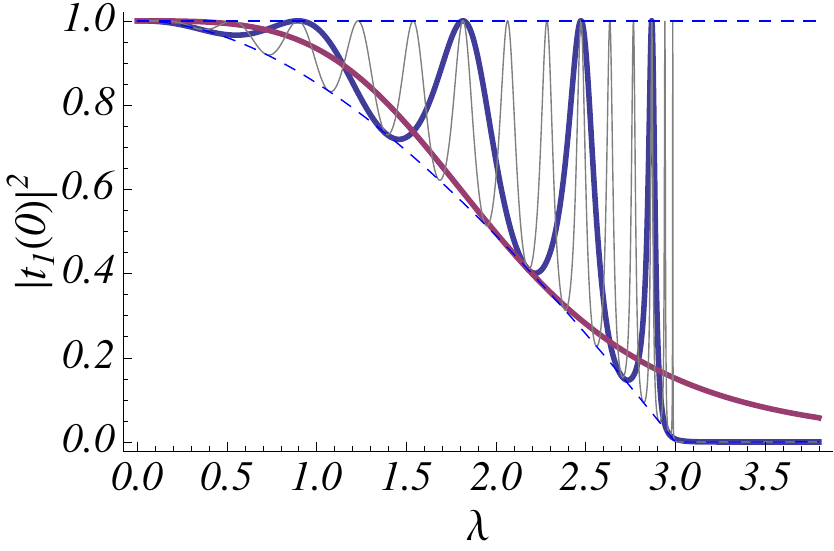}
\caption{Transmission probability $|t_1(0)|^2$ for a normally incident
quasiparticle of the first lower band with energy $E=2$ trought a barrier
of height $V_0=5$ (both in unit of $\hbar v/\ell$), with momentum 
$k_x^{(1)}$, Eq.~(\ref{k1}),
as a function of: 
(Left) the width of the barrier $d$ (in unit of $\ell$), 
for different values of $\lambda$ (in unit of $\hbar v/\ell$): 
$\lambda=0$ (gray line), $\lambda=1.2$ (red dotted line), $\lambda=2.8$ 
(yellow dashed line), $\lambda=4$ (blue solid line); 
(Right) the strength of the coupling $\lambda$, for different values of $d$: 
$d=2$ (red line), $d=5$ (blue line), $d=15$ (gray line).
The dashed lines are the boundaries determined by Eq.~(\ref{boundary1}).
}
\label{fig.sec1}
\end{figure}

Since Klein tunneling in graphene (and in general for relativistic particles) 
is expected to occur for normal incidence, 
let us consider in particular the transmission for $\phi=0$ and $V_0>E$. 
In this case the transmission coefficient takes a simple analytic 
expression which is the following
\begin{equation}
\label{t1_0}
t_1(\phi=0)=
\frac{e^{-i \ene_E d} \,\ene_E \,\ene_{E-V_0}}
{\ene_E \,\ene_{E-V_0} \,\cos\left(
\ene_{E-V_0} d\right)-i \big(\ene_E^2-(E+\lambda/2) V_0/v^2\big)\, 
\sin\left(\ene_{E-V_0} d\right)}
\end{equation}
The transmission probability $|t_1(0)|^2$ for normally incident particles coming from 
the first lower band is shown in Fig.~\ref{fig.cont1} and Fig.~\ref{fig.sec1}. 
As one can see, there is a strong suppression of the transmission for $\lambda>V_0-E$ 
while, for $0<\lambda<V_0-E$, $|t_1(0)|^2$ can oscillate between its maximum value, $|t_1(0)|^2=1$, occuring when $\ene_{E-V_0}d=n\pi $ (with $n=0,1,2,\dots$), 
and a lower bound which depends on $\lambda$, $E$ and $V_0$, occurring for 
$\ene_{E-V_0}d=n\pi /2$ 
(the lower envelope dashed curve drawn in Fig.~\ref{fig.sec1} (right panel)).\\
To conclude, the transmission probability, 
for $0<\lambda<V_0-E$, takes values in the following range
\begin{equation}
\label{boundary1}
\frac{\ene_E^2 \,\ene_{E-V_0}^2}{\big(\ene_E^2-(E+\lambda/2) V_0/v^2\big)^2}
\le |t_1(0)|^2\le 1,
\end{equation}
as shown in Fig.~\ref{fig.sec1}. In particular, the ridges in Fig.~\ref{fig.cont1}, 
corresponding to perfect transmission, $|t_1(0)|^2=1$,   
are described by the equations 
$d=\frac{n\pi v}{\sqrt{(E-V_0)(E-V_0+\lambda)}}$, with $n=0,1,2,\dots$ 
\textcolor{black}{Notice that the normal transmission probability $|t_1(0)|^2$ is an oscillating function also of the potential barrier $V_0$, and, even for $V_0$ very large, $V_0\rightarrow \infty$, it keeps on oscillating between ${v^2\ene_E^2}/{\big(E+\lambda/2\big)^2}$ and $1$.} 
%
\subsubsection{Upper band}
Let us now consider an incident particle carrying momentum $k_x^{(2)}$, Eq.~(\ref{k2}). 
In this case the scattering problem can be formulated in terms of the 
following equation
\begin{equation}
\label{eq2band}
\begin{pmatrix}
0\\
r_1\\
1\\
r_2
\end{pmatrix}=
{\cal T}
\begin{pmatrix}
t_1\\
0\\
t_2\\
0
\end{pmatrix}
\end{equation}
where now the transmission coefficient of the incident particle is $t_2$. 
Defining
\begin{equation}
\eneq_E=\frac{1}{v}\sqrt{E(E-\lambda)}
\end{equation}
the momentum can be written in terms of the incident angle $\phi$
\begin{eqnarray}
k_x=\eneq_E \cos\phi\\
k_y=\eneq_E \sin\phi
\label{qkyphi}
\end{eqnarray}
As before, since $V(x)$ is only along $x$-direction, $k_y$ is the same 
everywhere, also inside region $II$, 
$k_y=\eneq_E \sin\phi=\eneq_{E-V_0} \sin\theta$. 
We remark that in order to have a traveling incident particle in region $I$, with momentum in the upper band, namely in order to have 
a well-defined scattering problem, the value of the coupling is restricted to  
\begin{equation}
\label{cond_band2}
\lambda< E.
\end{equation}
Using Eq.~(\ref{qkyphi}) in Eq.~(\ref{k2}), 
in order to rewrite Eqs.~(\ref{W}), (\ref{transferM}) in terms of the angle of incidence $\phi$, solving Eq.~(\ref{eq2band}), 
we get $t_2(\phi)$, the transmission coefficient for an incident 
particle carrying momentum $(k_x, k_y)=\eneq_E(\cos\phi,\sin\phi)$.  
In Fig.~\ref{transmission_angle} (right plot) examples of the transmission probability $|t_2(\phi)|^2$ for different values of $\lambda$ are reported.
\begin{figure}[h!]
\includegraphics[width=.37\textwidth]{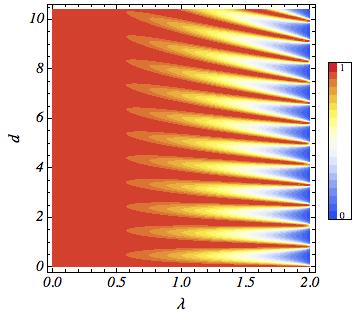}
\caption{Transmission probability $|t_2(0)|^2$, from Eq.~(\ref{t2_0}), 
for a normally incident quasiparticle coming from 
the second upper band 
with energy $E=2$ trought a barrier of height $V_0=5$ 
(both in unit of $\hbar v/\ell$),  with momentum
$k_x^{(2)}$, Eq.~(\ref{k2}),
as a function of the width of the barrier $d$ (in unit of $\ell$) 
and strength of the coupling
$\lambda$ (in unit of $\hbar v/\ell$) which is defined only for $\lambda<E$.}
\label{fig.cont2}
\end{figure}
\begin{figure}[h!]
\includegraphics[width=.37\textwidth]{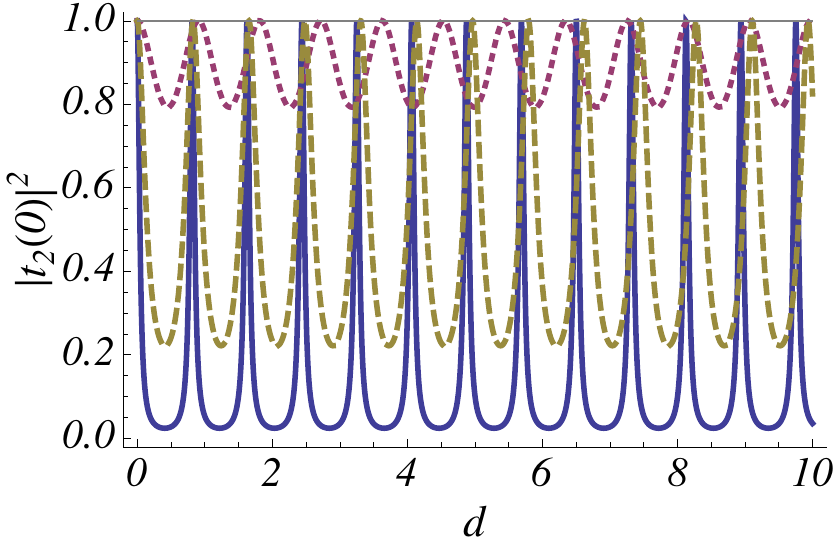}\hspace{0.2cm}
\includegraphics[width=.37\textwidth]{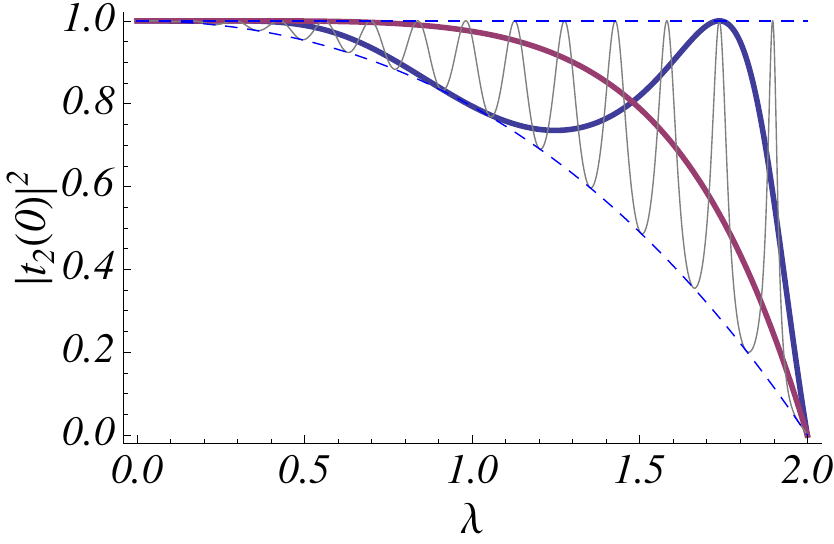}
\caption{Transmission probability $|t_2(0)|^2$ for a normally incident
quasiparticle of the first lower band with energy $E=2$ trought a barrier
of height $V_0=5$ (both in unit of $\hbar v/\ell$), with momentum
$k_x^{(2)}$, Eq.~(\ref{k2}),
as a function of: 
(Left) the width of the barrier $d$ (in unit of $\ell$), 
for different values of $\lambda$ (in unit of $\hbar v/\ell$):
$\lambda=0$ (gray line), $\lambda=1$ (red dotted line), $\lambda=1.8$ 
(yellow dashed line), $\lambda=1.98$ (blue solid line); 
(Right) the strength of the coupling $\lambda$, for different values of $d$:
$d=2$ (red line), $d=5$ (blue line), $d=50$ (gray line). 
The dashed lines are the boundaries given by Eq.~(\ref{boundary2}).
}
\label{fig.sec2}
\end{figure}

Considering particularly the transmission for $\phi=0$ and $V_0>E$, 
the coefficient $t_2(0)$ takes a simple analytic 
expression, analogous to Eq.~(\ref{t1_0}) where $\lambda\rightarrow -\lambda$, 
\begin{equation}
\label{t2_0}
t_2(\phi=0)=
\frac{e^{-i \eneq_E d} \,\eneq_E \,\eneq_{E-V_0}}
{\eneq_E \,\eneq_{E-V_0} \,\cos\left(
\eneq_{E-V_0} d\right)-i \big(\eneq_E^2-(E-\lambda/2) V_0/v^2\big)\, 
\sin\left(\eneq_{E-V_0} d\right)}
\end{equation}
Some results of Eq.~\ref{t2_0} are reported in Fig.~\ref{fig.cont2} and 
Fig.~\ref{fig.sec2}.
As already said, the permitted values for $\lambda$ are delimited 
by Eq.~(\ref{cond_band2}). 
From Fig.~\ref{fig.sec2} (left panel) the oscillations of the 
transmission resemble those for a non-chiral particle \cite{katsnelson2006}.  
For $0<\lambda<E$ the transmission $|t_2(0)|^2$ can vary
between its maximum value, $|t_2(0)|^2=1$ when $\eneq_{E-V_0}d=\pi n$ (with $n=0,1,2,\dots$) and a lower bound which 
depends on $\lambda$, $E$ and $V_0$, occurring when $\eneq_{E-V_0}d=\pi n/2$ 
(the lower envelope dashed line drawn in Fig.~\ref{fig.sec2} (right panel)).
The values of the transmission probability, therefore, 
take values in the following range
\begin{equation}
\label{boundary2}
\frac{\eneq_E^2 \,\eneq_{E-V_0}^2}{\big(\eneq_E^2-(E-\lambda/2) V_0/v^2\big)^2}\le
|t_2(0)|^2\le 1,
\end{equation}
as shown in Fig.~\ref{fig.sec2}.
\textcolor{black}{The normal transmission probability $|t_2(0)|^2$ is also an oscillating function of the potential barrier $V_0$ and, for $V_0\rightarrow \infty$, it oscillates between ${v^2\eneq_E^2}/{\big(E-\lambda/2\big)^2}$ and $1$.}

\bigskip
\textcolor{black}{In conclusion, for vanishing coupling $\lambda$ we recover the known result for massless chiral fermions exhibiting a perfect normal transmission through a potential barrier (Klein tunneling) due to conservation of pseudospin, namely the particle wavefunction at the barrier interface matches perfectly the hole wavefunction with the same pseudospin. 
On the contrary, for large $\lambda$ we get anti-Klein tunneling, known to occur for chiral fermions with effective mass, due to particles transforming into holes inside the barrier carrying imaginary momenta, so described by evanescent waves. \\
Interestingly, in the intermediate regime, dealing with the complete four-bands description where both pseudospin and spin (for the Rashba case) or interlayer-pseudospin (for the bilayer case) are present, we have oscillations and resonances characterized by perfect normal transmission for some particular values of the parameters, a behavior which resembles that of non-chiral particles in gapless semiconductors \cite{katsnelson2006}.} 


\section{Conclusions}
We studied a scattering problem in a single layer graphene in 
the presence of Rashba spin-orbit coupling. 
This system has the same transport properties of bilayer graphene where the interlayer coupling plays the role of the Rashba coupling in the monolayer case. 
We calculated the transmission probability through a potential barrier 
showing that, by increasing the coupling one can go from Klein tunneling regime,  
exhibited by free monolayer graphene, to anti-Klein tunneling, 
as expected in the bilayer graphene with strong interlayer coupling. 
We show that in the intermediate regime the transmission can develop 
resonances where perfect transmission is allowed for some particular values of the coupling or width of the barrier. We found, therefore, that 
the crossover from Klein to anti-Klein tunneling 
is not described by a simple monotonically decrease of the transmission 
amplitude but has a richer behavior. We are sure that our findings can be of 
some relevance in the view of very recent experiments on Klein tunneling 
in bilayer graphene \cite{arxiv}.

\makeatletter 
\makeatother


\end{document}